\documentclass[preprint]{aastex}
\usepackage{graphicx}
\usepackage{amsmath}

\shorttitle{Merger Origin of the Milky Way's Thick Disk}
\shortauthors{Dierickx et al.}

\begin{document}

\title{Observational Evidence from SDSS for a Merger Origin of the Milky Way's Thick Disk}



\author{Marion Dierickx\altaffilmark{1,2}, Rainer J. Klement\altaffilmark{2}, Hans-Walter Rix\altaffilmark{2} \& Chao Liu\altaffilmark{2}}

\altaffiltext{1}{Harvard University}
\altaffiltext{2}{Max-Planck-Institut f\"ur Astronomie, Heidelberg}

\begin{abstract} 
We test competing models that aim at explaining the nature of stars in the Milky Way
that are well away ( $\vert z\vert \gtrsim 1$ kpc) from the midplane, the so-called thick disk: the
stars may have gotten there through orbital migration, through satellite
mergers and accretion or through heating of pre-existing thin disk stars.
\citet{sales09} proposed the eccentricity distribution of thick disk stars as a diagnostic
to differentiate between these mechanisms. Drawing on SDSS DR7, we have
assembled a sample of 34,223 G-dwarfs with 6-D phase-space information and
metallicities, and have derived orbital eccentricities for them.
Comparing the resulting eccentricity distributions, $p(e|z)$,  with the
models, we find that: a) the observed $p(e|z)$ is inconsistent with that
predicted by orbital migration only, as there are more observed stars of high
and of very low eccentricity; b) scenarios where the thick disk is made
predominantly through abrupt heating of a pre-existing thin disk are also inconsistent,
as they predict more high-eccentricity stars than observed; c) the
observed $p(e|z)$ fits well with a ``gas-rich merger'' scenario, where most thick disk stars were born from unsettled gas \emph{in situ}.
\end{abstract}

\section{Introduction}
\label{sec:intro}
The common presence of so-called 'thick' stellar disk components in spiral
galaxies, including in our Milky Way, has been established for some decades
\citep{bur78,gil83,yoa06}: the vertical distribution of stellar brightness or mass is more adequately fit with a
thin -- thick disk superposition, rather than with a single (thin) disk 
component, where the thick disk has a scale height that
is typically three times larger. Thick disk stars further differ from stellar populations closer to
the galactic midplane in terms of age, metallicity and rotational velocity. 
Certainly in the Milky Way, thick disk stars are generally older, more metal-poor, and
rotate more slowly around the galactic center \citep[e.g.][]{feltz08,reddy09}. 

There is no consensus on the origin of the thick disk. Qualitatively
different scenarios for the origin of thick disks have been discussed and cast in simulations, resulting in 
differing chemical and kinematic properties for the thick disk stellar
population. \citet{sales09} examined examples of these different scenarios and
proposed the orbital eccentricity distribution of stars at 1--3 thick disk scale heights
from the plane as a model discriminant for the case of the Milky Way. 
Broadly speaking, two of the four scenarios in \citet{sales09} presume that the thick disk stars 
got heated from a once thinner disk, and two explain the thick disk as a consequence
of material that was deposited in the course of a minor merger.

The first of these scenarios, {\it radial migration}, is based entirely on
internal processes, and the thick disk is created from stars migrating 
outwards from the kinematically hotter inner regions of the Milky Way 
\citep{roskar08,sch09}. In the second scenario, {\it heating},
the thick disk is predominantly the result of rapid heating of a pre-existing thin disk through one
relatively massive merger event \citep{villa08}. In this scenario,
a modest fraction of the original thin disk is 
preserved \citep{kazan08,villa08}, and a small portion of present-day thick
disk stars were formed in the satellite galaxy and are on highly eccentric orbits.

In the {\it accretion} scenario, thick disk stars mostly form in an external
satellite galaxy that gets disrupted while on a prograde orbit near the
disk plane. This can produce many properties of observed old thick disk 
components \citep{abadi03}. Finally, the {\it gas-rich merger} scenario is 
a variant of this idea, where a minor merger epoch brings gas into the galaxy
from which (thick disk) stars form before the gas completely settles 
into a thin disk.

Analyzing four concrete models based on these scenarios,
\citet{sales09} proposed orbital eccentricity
as a comparative diagnostic. For each simulation, they made eccentricity
distributions for simulated stars distant enough from the midplane 
to eliminate thin disk stars. They pointed out that the eccentricity distribution
$p(e\vert z)$ differed significantly between these four scenarios (reproduced
as histograms in their Fig.~3). 

The aim of this {\it Letter} is to compare these predictions to the observed
eccentricity distribution of thick disk stars in the Milky Way,
drawing on G-dwarfs from SEGUE DR7 with 3D position and 3D velocity information.
At face value, this comparison favors the gas-rich merger scenario as the 
most important mechanism of thick disk formation. We also point out that the relationship between height, metallicity and
eccentricity of thick disk stars has great potential for more stringent data--model
comparisons with existing data.

\section{Data}
\label{sec:data}

\subsection{{\it Sample Definition}}
\label{sec:data selection}
\citet{sales09} calculated their diagnostic eccentricity distribution in ``solar neighbourhood volumes'',
which they devised as cylindrical shells between two and three scale radii of the model galaxies' thin disk ($2<R/R_\text{d}<3$).
To reject most of the thin disk stars, they further constrained the stars to be at $1< \vert z_\text{scaled}\vert <3$ ($z_\text{scaled}\equiv z/z_0$).
For the vertical thick disk scale height $z_0$,
\citet{sales09} used $\sim 1$kpc, except for their {\it accretion} model, which had $z_0=2.3$ kpc.

It is impossible to match such a volume with any set of actual observations. Here, we draw 
observations mostly from the conical volume (with the Sun at its apex) that is naturally
provided by the SDSS North Galactic Cap survey area. Specifically, we drew up a 
sample of 34,223 G-dwarf candidate stars from SDSS DR7 \citep{aba09}, for which spectra were taken
and spectral parameters had been devised. Our selection criteria are outlined in Table~\ref{tab:t1},
and we provide the reader with the full data in Table~\ref{tab:t2} (available electronically only).
G dwarfs are the best-sampled spectroscopic target category in SEGUE \citep{yan09}, and 
a good fraction of these spectroscopic targets falls into the distance range of 0.8-6 kpc that 
is of particular interest here.  

Photometric distances (and errors) for these target stars were derived based on their $ugr$ colors and 
apparent magnitudes, following the prescription of \citet{ive08} which we expect to yield systematic errors $\lesssim5\%$ \citep{kle09}.
These heliocentric distances allow us to derive the distribution of heights above the plane for these stars. 
For the model comparison, we need a sample of stars whose distribution 
$p(z)$ reasonably matches $exp(-z/z_0)$, with $z_0\sim1$ kpc \citep[as in][]{sales09}. 
We found that in the range $1 < \vert z_\text{scaled}\vert < 3$, the $p(z)$ of our sample matches
$exp(-z/z_0)$ to within 20\%; 
apparently, the SDSS selection function, the G dwarf luminosity distribution 
and the conical volume compensate to provide us with the appropriate height distribution. 


\begin{deluxetable}{cc}
\tablewidth{0pt}
\tablecolumns{2}
\tablecaption{Selection criteria for SEGUE data.}
\tablehead{
\colhead{Parameter} & \colhead{Criterion}
}
\startdata
$(g-r)_{0}$ & 0.48 $<(g-r)_{0}<$ 0.55 \\
$(u-g)_{0}$ & 0.6 $< (u-g)_{0}<$ 2.0 \\
$(r-i)_{0}$ & -0.1 $<(r-i)_{0}<$ 0.4 \\
Absolute velocities & $|v| < $ 600 km/s for $v_{x}$, $v_{y}$, $v_{z}$\\
$T_{eff}, [Fe/H], [\alpha/Fe]$ & have valid values \\
$log(g)$ & $log(g) > 3.75$ \\
$E(B-V)$ & $E(B-V) < $ 0.3 \\
\enddata
\tablecomments{The three first rows impose color restrictions valid for G stars.
  The fourth removes stars that are gravitationally unbound. Row five
  specifies the need for $T_{eff}, [Fe/H], [\alpha/Fe]$ data. Row 6 imposes
  high $g$ to select dwarfs for which photometric distances can be estimated.
  The bottom row minimizes interstellar extinction.}
\label{tab:t1}
\end{deluxetable}

\begin{deluxetable}{cc}
\tablewidth{0.45\textwidth}
\tablecolumns{2}
\tablecaption{Observed and derived properties of our 34,223 SEGUE stars.}
\tablehead{
}
\startdata
\enddata
\tablecomments{Available online as a machine-readable table.}
\label{tab:t2}
\end{deluxetable}
\subsection{{\it Phase-space coordinates}\label{sec:coor and vel}}

The directly observed quantities for each star are its position, photometry,
proper motion and its spectroscopic parameters, with the 
line-of-sight velocity and the metallicity [Fe/H] most relevant here.
These quantities need to be translated into the phase-space coordinates
$\vec{\mathbf r}$ and $\vec{\mathbf v}$ and their errors in the Galactocentric
reference frame. This is done using conversion matrices from \citet{john87},
with $v_{c}=220$ km.s$^{-1}$  for the circular velocity of the Local Standard of Rest (LSR) 
and 8~kpc for the Sun's distance to the Galactic Center. The errors for each 
$\vec{\mathbf v}_i$ were derived following \citet{john87}, using the 
radial velocity and proper motions provided in DR7 and distance measurements
taken from above. Typical errors are $\sim 5$ km.s$^{-1}$ for the line-of-sight 
component \citep{yan09}\footnote{Our formal radial velocity errors obtained from the database are smaller by a factor of $\sim2$;
however, this only has a minor effect on the spatial velocity components compared to the proper motion errors.},
and $\sim 25$ km.s$^{-1}$ for each transverse component at a distance of 2 kpc.
The errors for each galactocentric $\vec{\mathbf r}_i$ are 
calculated simply from the distance error under
the assumption that stellar galactic longitude and latitude are known precisely.
The importance of both the individual uncertainties and of possible
systematic uncertainties in the distance scale or the circular velocity at the 
solar radius are assessed through their impact on the derived eccentricity 
distribution in the next section.

\section{Analysis}
\label{sec:analysis}

\subsection{Eccentricity Estimates\label{sec:ecc}}
In order to calculate the orbital eccentricities for each star, we need to adopt
a gravitational potential. Here we chose a simple logarithmic potential,
$\Phi = v_{c}^2 * \ln r $, where $r$ is the distance from the Galactic Center 
to the star in spherical coordinates. With this potential, we can calculate 
for each star $i$ its total angular momentum $L_i$, its energy $E_i$, and the effective potential:
\begin{equation}\label{eq:e1}
\Phi_\text{eff}(r_{i}) = \Phi(r_{i}) + \frac{L_{i}^2}{2r_{i}^2}\,.
\end{equation}
At the peri- and apocenter of the orbit of star $i$, the energy equals the effective potential:
\begin{equation}\label{eq:e2}
E_{i} = \Phi_\text{eff} (r_\text{apo/peri})\,, 
\end{equation}
where we solved for the two roots $r_\text{apo/peri}$ in
equation \eqref{eq:e2} by simple bisection.
The eccentricity is then defined and given as
\begin{equation}
e = \frac{r_\text{apo} - r_\text{peri}}{r_\text{apo}+r_\text{peri}}\,.
\end{equation}\\

The resulting distributions of observed orbital eccentricities 
are shown -- as red histograms -- in Fig.~\ref{fig:f1}, taken over height
ranges $1< \vert z_\text{scaled}\vert<3$. 

\subsection{Eccentricity Uncertainties}
Before discussing these distributions,
we assess the impact of two important error sources on these
distributions: the individual phase-space uncertainties and the choice
of the potential. For each stellar $(\vec{\mathbf r},\vec{\mathbf v})_i$ with
its associated $(\delta\vec{\mathbf r},\delta\vec{\mathbf v})_i$, we created
100 Monte-Carlo realizations of phase space coordinates, calculated
the eccentricities and created 100 realizations of the eccentricity histogram.
These showed that the individual measurement uncertainties result in
little variance in the eccentricity histogram and also cause no significant bias
(e.g. away from small eccentricities). We also explored what the impact of
a 5\% systematic uncertainty in the distance (from the Sun) would be and
found the effect on $p(e|z)$ to be small.
We also compared $p(e|z)$ resulting from our choice of the simplistic
logarithmic potential to the distribution produced by a more complex
(and realistic) potential, consisting of a logarithmic halo, a Hernquist
bulge and a Miyamoto-Nagai disk \citep[e.g.][]{joh99}.  
The one-to-one correlation of the resulting eccentricities showed very
little scatter for eccentricities less then 0.6. For higher eccentricities,
the two estimates still show a good correlation, but with some net bias to
larger eccentricities in the three-component potential. Adopting this
more complex form for the potential would leave $p(e|z)$ largely
unaffected for $e<0.6$ and slightly boost the high eccentricity tail.
We also explored the impact of changing $v_{c}$ by $\pm$
10 km.s$^{-1}$ in the logarithmic potential context and found it to have no 
significant influence on the eccentricity distribution. This is because
$v_{c}$ enters both into the transformation to the Galactic rest-frame and
into the gravitational potential.

\subsection{Cutting the Sample}
To focus on the eccentricity distribution of thick disk stars and to minimize the contribution
from the halo, 
we only consider stars with tangential velocity $v_{\Phi}$ greater than 50 km.s$^{-1}$ in the
Galactocentric rest-frame system \citep[i.e. we remove all stars on slowly-rotating and retrograde orbits, consistent with][]{sales09}.
The effect of this cutoff on the overall shape 
of the eccentricity distribution is negligible. For the histograms in Fig.~\ref{fig:f1},
we also eliminated the small fraction of stars with metallicity $\text{[Fe/H]} \leq -1.2$, 
as they may be chemically attributable to the halo. This does not affect the overall 
result except for somewhat reducing the incidence of very high
eccentricities ($e \geq 0.6$).  We return to the question of the correlation between 
kinematics and metallicity at the end, though it was not explicitly considered 
by \citet{sales09}. Together, these restrictions remove 10.6\% of stars in the original
data set.\\

\section{Results}
\label{sec:results}

\begin{figure*}[bt]
\begin{center}
\includegraphics[scale=0.65]{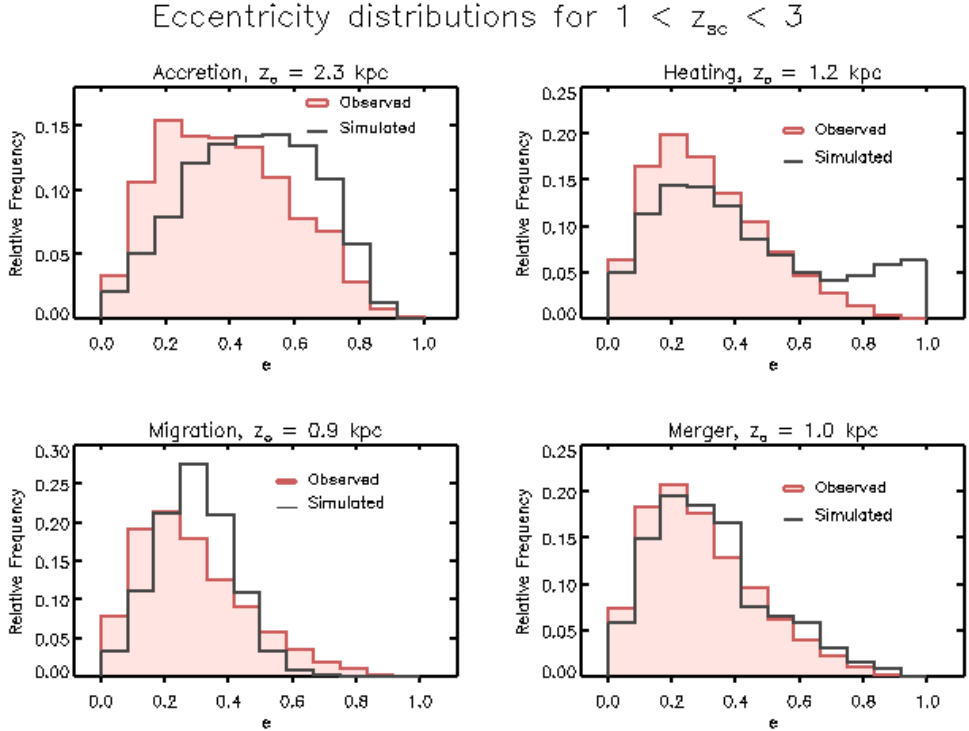} 
\caption{Eccentricity distributions at a given height range above the
  midplane. \label{fig:f1}}
\end{center}
\end{figure*}

We now turn our attention to the main result of this {\it Letter}, the comparison
of the observed eccentricity distributions derived from SDSS/SEGUE
G dwarfs with the model predictions by \citet{sales09}; this comparison is
summarized in Fig.~\ref{fig:f1}, where the panels are in the same order as in \citet[][their Fig.~3]{sales09}.
The red histograms show the observed $p(e|z)$ for the
four different scenarios, each scaled to $z_\text{0}$ \citep[Table~1 from][]{sales09}.
For all but the {\it accretion} scenario (top left panel) these distributions
look very similar, with a peak at $e\approx 0.25$ and a pronounced 
asymmetric tail towards high eccentricities, $e\approx 0.9$. 
Unsurprisingly, for the accretion scenario the mode of the distribution is shifted towards higher eccentricities, 
as we sample the kinematics from 2 kpc $\lesssim z \lesssim 7$ kpc from the 
disk midplane.

Figure~\ref{fig:f1} shows that neither one of the two processes where the thick disk consists
predominantly of stars that were ``kicked up'' from a once thinner configuration matches
the observed $p(e|z)$. Orbital migration (bottom left) predicts an eccentricity 
distribution that is too narrow. The data show a higher fraction of stars both on nearly circular
and on highly eccentric orbits. The heating scenario (top right), where much of the
thick disk was puffed up from a thinner disk through a massive merger, provides a 
reasonable match to the observations for $e\le 0.6$. However, this scenario also predicts 
that the stars from the satellite involved in this merger should be part of the 
thick disk, but at very high eccentricities $0.7<e<1$; the data show now evidence for this. 
In the accretion scenario (top left) much of the thick disk consists of deposited
satellite debris; as this panels shows, this predicts too high eccentricities even for the bulk of
thick disk stars. 

The predictions of the gas-rich merger scenario (bottom left) match the observations well, at
least qualitatively if not in the formal sense: there is a maximum in $p(e|z)$ at 0.25 and a tail extending
to $e=0.9$. Recall that in this scenario both the pre-existing disk and the merging satellite are very 
gas-rich. Therefore, e.g. the ``culprit'' stars from the satellite which imply many high-eccentricity stars in
the heating scenario (top right panel of Fig.~\ref{fig:f1}) are absent. 

\section{Discussion and Conclusions}
\label{sec:discussion}
We have compared the predictions made by four thick disk formation simulations
to eccentricity distributions at $\vert z\vert \gtrsim1$ kpc resulting from SEGUE data. Eccentricity proves a
valuable parameter as it allows to clearly differentiate and assess the four
scenarios outlined by \citet{sales09}. Direct accretion of
stars from satellites generally predicts orbits that are too eccentric. Radial
migration fails to generate the observed number of very low and very high
eccentricity stars. Kinematic heating of a primordial thin disk, on the other
hand, predicts too many stars with highly eccentric orbits accreted from the
disrupted satellite. The accretion of gas onto the disk as a result of
satellite mergers is generally consistent with the observed
distribution.

Taken at face value, the comparison of the eccentricity distributions argues strongly against
three of the four scenarios; they favor an origin of the thick disk ($\vert z\vert\gtrsim1$ kpc) from a series of gas-rich (minor) mergers,
where most thick disk stars formed {\it in situ} and a high eccentricity tail arises from accreted stars.
However, the simulations underlying three of the four scenarios were aimed
at elucidating the physical processes capable of producing a thick disk component in late-type spirals; they were
not necessarily tuned to match Milky Way properties 
\citep{brook04,abadi03}.  Therefore, they may not represent the full range of 
possible $p(e|z)$ within these scenarios.  Also, these scenarios are of course not mutually 
exclusive. For example, orbital migration must be present at some level,
irrespective of the minor-merger history of the Milky Way. The bottom left panel of Fig.~\ref{fig:f1}
only argues against orbital migration as the sole or dominant process.\\
 
\begin{figure*}[bt]
\begin{center}
\includegraphics[scale=0.4]{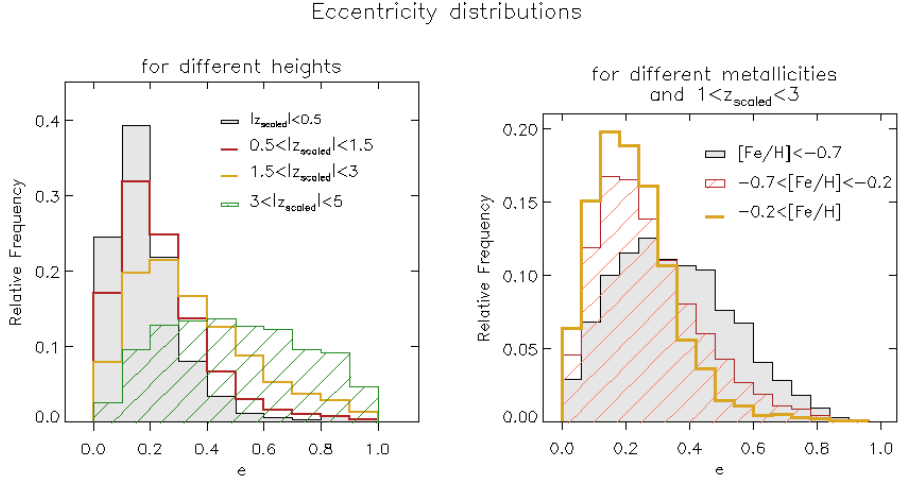} 
\caption{Eccentricity distributions for a range of heights and metallicities. \label{fig:f2}}
\end{center}
\end{figure*}

Speculation about variants of the seemingly rejected scenarios that could match the data
is beyond the scope of this {\it Letter}. However, we want to point out the wealth of 
chemo-kinematic information that is now available to further test the origin of the thick disk. 
In Fig.~\ref{fig:f2} we use the G-dwarf sample to illustrate how well the distribution of
stellar orbits within the thick disk can be mapped as a function of height as well
as metallicity: as expected, orbits become more eccentric as one moves away from the
plane. At a given height above the plane, the most metal-poor stars are on the 
most eccentric orbits, while stars on nearly
circular orbits have metallicities closer to that of the Sun \citep[see also][]{ive08}.

Such information provides further tests of the scenarios. E.g., \citet{villa08} find
that in their thin disk heating simulation, the
fractional number of accreted stellar particles increases with height above
the galactic midplane.  And \citet{sales09} argue that regardless of the
specific formation mechanism considered, accreted stars are always associated
with the high eccentricity end of the distribution. 

Ultimately, the observed distributions, as e.g. in Fig.~\ref{fig:f2}, should be tested against 
further predictions made by simulations in order to substantiate or refute gas-rich satellite
mergers as a plausible thick disk formation mechanism in the Milky Way.

\acknowledgments

Funding for the SDSS and SDSS-II has been provided by the Alfred P. Sloan
Foundation, the Participating Institutions, the National Science
Foundation, the U.S. Department of Energy, the National Aeronautics and
Space Administration, the Japanese Monbukagakusho, the Max Planck Society,
and the Higher Education Funding Council for England. The SDSS Web Site is
http://www.sdss.org/.\\
MD acknowledges support from the Weissman International Internship Program which enabled an
extended visit to MPIA, where much of this work was carried out.

\clearpage

\end{document}